\documentclass[twocolumn,secnumarabic,amssymb, nobibnotes, aps, prd,superscriptaddress]{revtex4-1}

\usepackage{longtable}
\usepackage{graphicx}
\usepackage{amsmath,amssymb}
\usepackage{textcomp}
\usepackage[normalem]{ulem}
\usepackage{todonotes}
\usepackage{enumitem}   

\begin{document}

\title{Breakup mechanisms in the $^{6}$He+$^{64}$Zn reaction at near-barrier energies}%

\author{J. P. Fern{\'a}ndez-Garc{\'i}a }%
\email[]{jpfernandez@us.es}
\affiliation{Departamento de FAMN, Universidad de Sevilla, Apartado 1065, E-41080 Seville, Spain}
\affiliation{Centro Nacional de Aceleradores, Universidad de Sevilla, Junta de Andaluc{\'i}a-CSIC, 41092 Sevilla, Spain}

\author{A. Di Pietro}
\affiliation{INFN, Laboratori Nazionali del Sud, via S. Sofia 62, 1-95123 Catania, Italy}

\author{P. Figuera}
\affiliation{INFN, Laboratori Nazionali del Sud, via S. Sofia 62, 1-95123 Catania, Italy}


\author{J. G{\'o}mez-Camacho}
\affiliation{Departamento de FAMN, Universidad de Sevilla, Apartado 1065, E-41080 Seville, Spain}
\affiliation{Centro Nacional de Aceleradores, Universidad de Sevilla, Junta de Andaluc{\'i}a-CSIC, 41092 Sevilla, Spain}

\author{M. Lattuada}
\affiliation{INFN, Laboratori Nazionali del Sud, via S. Sofia 62, 1-95123 Catania, Italy}
\affiliation{Dipartamento di Fisica e Astronomia, via S. Sofia 64, I-95123 Catania, Italy}

\author{J. Lei}
\affiliation{Institute of Nuclear and Particle Physics, and Department of Physics and Astronomy,
Ohio University, Athens, Ohio 45701, USA}

\author{A. M. Moro}
\affiliation{Departamento de FAMN, Universidad de Sevilla, Apartado 1065, E-41080 Seville, Spain}

\author{M Rodr{\'i}guez-Gallardo}
\affiliation{Departamento de FAMN, Universidad de Sevilla, Apartado 1065, E-41080 Seville, Spain}

\author{V. Scuderi}
\affiliation{INFN, Laboratori Nazionali del Sud, via S. Sofia 62, 1-95123 Catania, Italy}

\date{\today}%
\begin{abstract}

New experimental results for the elastic scattering of $^{6}$He on $^{64}$Zn at incident energies of 15.0 and  18.0 MeV and $^4$He at 17.5 MeV along with results already published at 10.0 and 13.6 MeV, are presented. Elastic and $\alpha$ experimental cross sections are compared with coupled-reaction-channel, continuum-discretized coupled-channel and a DWBA inclusive-breakup models. The large yield of $\alpha$ particles observed at all measured energies  can be explained by considering a non-elastic breakup mechanism.



\end{abstract} 
\maketitle

\section{Introduction}
The understanding of peripheral heavy-ion collision processes in general, and elastic scattering in particular, is an important part of the overall understanding of heavy-ion reaction dynamics and its dependence on the structure of the colliding nuclei. Indeed, most reaction theories require as a prerequisite for their application the knowledge of the optical potentials derived from the elastic scattering of the particles involved. Despite being a peripheral process, elastic scattering shows direct evidence of the internal structure of the colliding nuclei; one example is given by the elastic scattering involving halo nuclei as projectiles. Many experimental and theoretical studies of scattering involving halo nuclei on various target masses have been performed so far and complete reviews can be found in \cite{Can06,Keeley2009,Kolata2016}. The dynamics of the elastic scattering process has shown many features related to the peculiar characteristics of these nuclei. The results of these studies can be summarized as follows: the low binding energy of the halo nuclei enhances the probability of breakup; as a consequence,  reduced elastic scattering and  large total reaction cross sections, with respect to the collision induced by the well bound core nucleus on the same target and  $E_{\rm c.m.}$, are found for all the investigated systems.

Coupling to nuclear and Coulomb breakup plays a relevant role; it modifies the elastic cross section  especially in the region of interference between the nuclear and Coulomb amplitudes, resulting in a damped Coulomb-nuclear interference peak. The size of the effect depends upon the B(E1) strength near the threshold of the halo projectile and upon the charge of the target \cite{San08,DiPietro10,Dipietro12,cubero11li}.

Coupling to transfer has also been found to influence elastic scattering. The role of coupling of one-neutron transfer to bound states of the target nucleus  has been discussed in \cite{Keeley2009} and it was found to be significant;  the  effect  depends on the detailed nuclear structure of the target nucleus. In the case of $^6$He-induced collisions the relevance of the two-neutron ($2n$) transfer channel has been  observed  \cite{Riccardo2004,DiPietro2004}. In \cite{Fer10plb,Esc07} the effect of two-neutron transfer channels on elastic and $\alpha$-particle production cross-section was investigated for the reaction $^6$He+$^{206}$Pb and found to be important. The coupling with these channels produce a strong effect on the elastic cross section, giving good agreement with the data. Moreover, it explains the energy and angular distribution of the $ \alpha $-particles produced in this reaction.

In  reactions induced by $^6$He, all the aforementioned effects seem to be equally important and need to be considered if a full account of the experimental results is sought.  In the present paper new results of elastic scattering data for the $^6$He+$^{64}$Zn reaction at $E_{\rm beam}$= 15.0 and 18.0 MeV measured with high accuracy are reported as well as the energy distribution of the $\alpha$-particles coming from the inclusive breakup of $^6$He at the same energies. 


In order to fully investigate and understand the various aspects of the reaction dynamics of the $^6$He two-neutron halo nucleus, the present experimental data, together with results of $^{4,6}$He+$^{64}$Zn  already published  \cite{DiPietro03,DiPietro2004,scuderi11}, have been described using different theories. This was necessary since, at present, one theory that is able to describe  all the experimental observables is not available.
Such a complete theoretical analysis has been performed within Optical Model (OM), Continuum-Discretized Coupled-Channel (3-body and 4-body CDCC) and Coupled-Reaction-Channel (CRC) formalisms. Moreover, the $\alpha$-particle spectra emitted in the reactions and reported in \cite{DiPietro2004,scuderi11} were theoretically analyzed considering, in addition to the elastic breakup contributions calculated with the CDCC method, the contribution of non-elastic breakup, computed according to the formalism proposed in \cite{Ich85}, and 2n transfer calculated with CRC. \

The paper is organized as follows: in  Sec.~\ref{sec:exp} the experiment is described. In Sec.~\ref{theor} the results of the different theoretical approaches used are reported. In Subsec.~\ref{sec:om} the $^4$He+$^{64}$Zn  and  $^6$He+$^{64}$Zn data are analyzed under the Optical Model (OM) framework. The subsequent paragraphs report on: Continuum-Discretized Coupled-Channel (Sec.~\ref{sec:cdcc}), Coupled-Reaction-Channel (Sec.~\ref{sec:crc}) and non-elastic breakup (Sec.~\ref{sec:nebu}) calculations for $^6$He+$^{64}$Zn. Finally, Sec.~\ref{sec:concl} is devoted to summary and conclusions.
\\


\section{Experimental data and set-up}\label{sec:exp}
\begin{figure}[h]
\begin{minipage}{20pc}
\includegraphics[width=20pc]{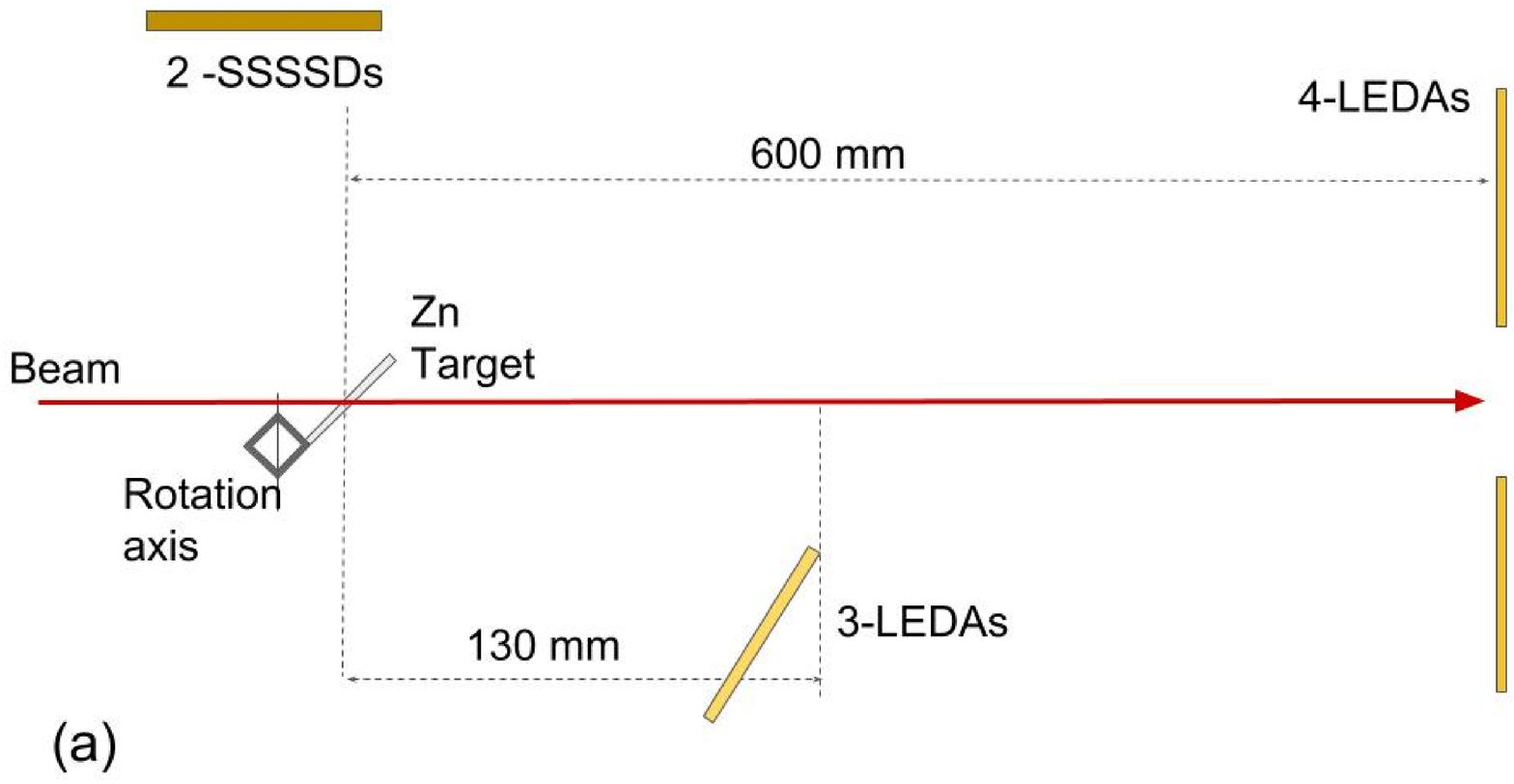}
\includegraphics[width=20pc]{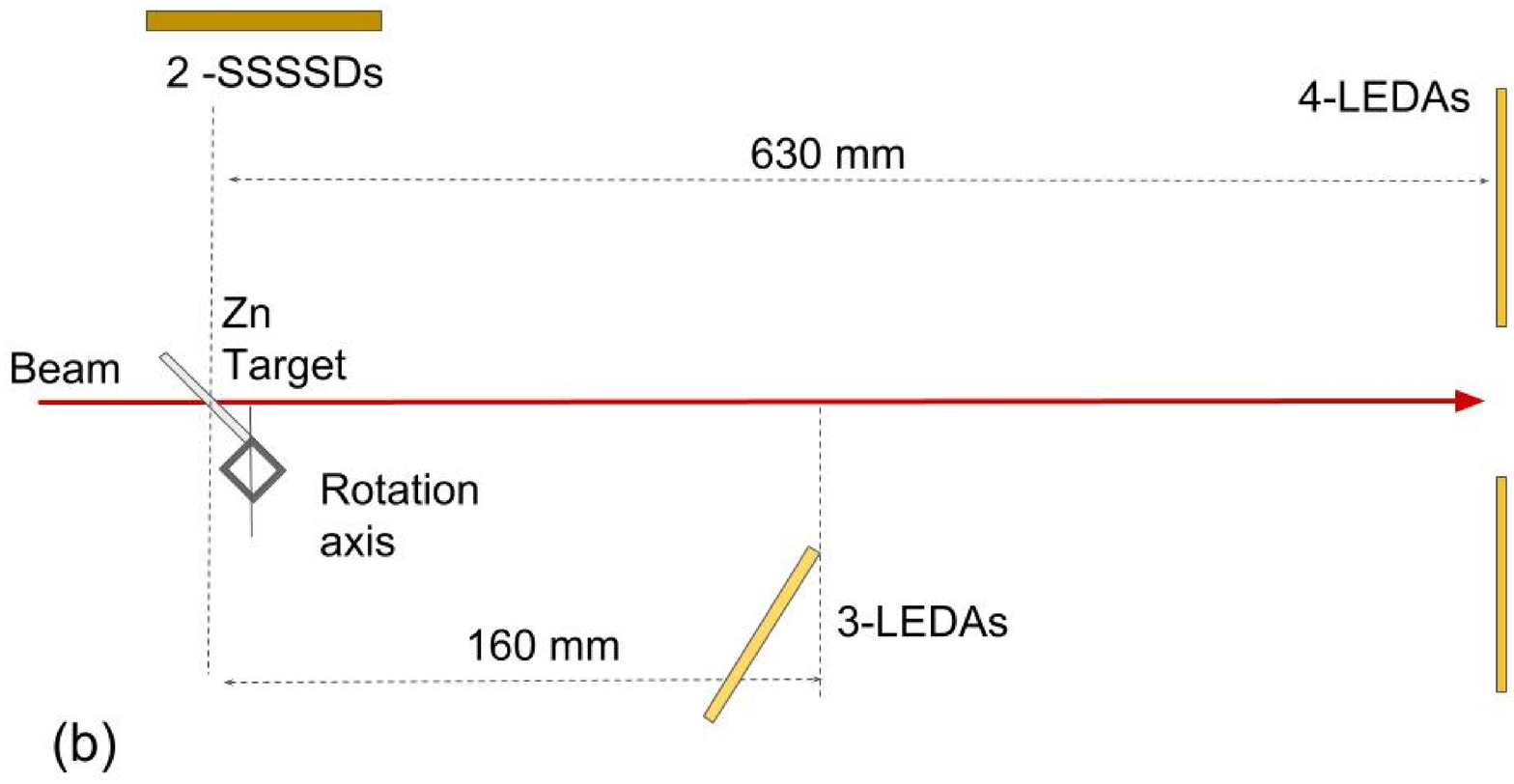}
\caption{\label{setup}Sketch of the experimental setup with the target tilted at +45$^{\circ}$ (a) and  -45$^{\circ}$ (b).}
\end{minipage}\hspace{2pc}%
\end{figure}

The experiments $^{4,6}$He+$^{64}$Zn were performed at the Centre de Recherches du Cyclotron at Louvain la Neuve (Belgium) in 2004. A radioactive beam of $^{6}$He at $E_{\rm beam}$ = 15.0 and 18.0 MeV with an average intensity of 3$\times$10$^{6}$ pps, and in addition a stable beam of $^{4}$He at  $E_{\rm lab}$= 17.5 MeV, were used. In these experiments the fusion excitation function as well as the $\alpha$-particle angular distributions  were measured and reported in a previous paper \cite{scuderi11}. In Fig.~\ref{setup}, a schematic view of the experimental set-up is shown. The target used was a 530 $\mu$g/cm$^2$ self-supporting $^{64}$Zn foil; it was tilted at $\pm$45$^\circ$ with respect to the y-axis (vertical direction) in order to allow measuring at laboratory angles around 90$^\circ$. Considering the target thickness, the energy at the center of the target was $E_{\rm lab}$ = 14.85 and 17.9 MeV for $^6$He and 17.4 MeV for $^4$He. Since the target was not rotating around its center, but around  its bottom-edge, by changing the target angle from +45$^\circ$ to -45$^\circ$, it was  possible to change the target-detector distance  and, as a consequence, the detector  angles. This allowed to measure the elastic scattering angular distribution in a wider angular range, and to reduce the angular gaps from one detector array to the other. 

Three arrays of silicon strip detectors consisting of seven sectors of LEDA-type \cite{Leda} and two Single-Sided Silicon Strip Detectors (SSSSDs) 50$\times$50 mm$^2$ were used.  The first array consisted of four LEDA sectors, 300 ${\mu}$m thick, placed in a symmetric configuration with each sector normal to the beam direction (up, down, left, right) at a distance of about 600 mm or 630 mm from the target (depending upon the target angle), covering the overall laboratory angular range 5$^\circ$ $\leq\theta\leq$ 12$^\circ$. Such a configuration allowed to monitor the beam misalignment and to normalize the cross-sections to the Rutherford scattering. Other three LEDA sectors, 500 ${\mu}$m thick, were angled at 45$^{\circ}$ with respect to the beam direction. They were placed close to the target (130 and 160 mm depending upon the target angle) and were covering an overall angular range of approximately 18$^\circ$ $\leq\theta\leq$ 67$^\circ$. This configuration allowed a very large solid angle coverage. The two SSSSDs detectors were placed around 90$^{\circ}$, parallel to the beam  axis, and covering the overall angular range 67$^\circ$ $\leq\theta\leq$ 120$^\circ$. The distance between the detector active area and the target was about 85 mm for both SSSSDs.  Helium was identified and clearly separated from Hydrogen by the ToF technique using as time reference the RF signal of the cyclotron. Time resolution was insufficient to separate different helium isotopes, therefore the Helium spectrum included $\alpha$-particles and elastically scattered $^6$He. \\

 Rutherford scattering on a Au target  along with a full  Monte Carlo simulation of the set-up was used to deduce the angle and solid angle of each detector strip. Once the geometry of the set-up was deduced for the two angular settings of the target, the full angular distributions of $^{4,6}$He+$^{64}$Zn were  obtained by normalizing the  elastic scattering data at smaller angles ($\theta{\le}10^\circ$) to the Rutherford cross-section, without further adjustment. \\ 
 A cross-check of the data published in \cite{DiPietro03,DiPietro2004} using the new Monte Carlo code was performed. By applying the new simulations to deduce the solid angles for the set-up used in \cite{DiPietro03,DiPietro2004},  a maximum difference of 5\% in the experimental angular distribution at $E_{\rm lab}$ = 13.5 MeV is found, with respect to those already published in \cite{DiPietro03,DiPietro2004}. This is within the experimental error bar reported in \cite{DiPietro03,DiPietro2004} which accounted for possible systematic errors in the determination of the cross-section.  In the data from \cite{DiPietro03,DiPietro2004} an additional systematic error in the absolute normalisation, of the order of 5\%, could be present as due to lack of data at very small angles where the cross-section is Rutherford.


\section{Theoretical calculations} \label{theor}

In this section we  compare the present experimental data as well as the one previously published in \cite{DiPietro03,DiPietro2004} with different theoretical calculations in order to understand the reaction mechanisms governing in the  $^{6}$He+$^{64}$Zn reaction at energies around the Coulomb barrier. First, we present an Optical Model (OM) analysis of the  $^{4}$He+$^{64}$Zn elastic cross sections. The potentials obtained from these calculations will be later used in the few-body calculations performed for the $^{6}$He+$^{64}$Zn reaction.

As a first  approach to analyze the $^{6}$He+$^{64}$Zn data,  OM calculations will be performed in Subsec.~\ref{sec:om}. Then, we will focus on the measured $\alpha$ cross sections. Since these data are inclusive with respect to the (unobserved) neutrons, they contain in general two distinct contributions, namely,   (i) the elastic breakup (EBU), in which the projectile fragments ``survive'' after the collision and the target remains in its ground state, and (ii) the non-elastic breakup (NEB), which includes any process in which the $\alpha$ particles ``survive'' whereas the dissociated neutrons interact non-elastically with the target nucleus (i.e. target excitation and neutron transfer or absorption). The EBU  can be taken into account by the CDCC calculations, while the CRC calculations account for EBU and NEB components approximately. Finally, an alternative calculation for the NEB part is presented, in which the closed-form method of Refs.~\cite{Ich85,Jin15} is employed. The same di-neutron model of $^{6}$He ($\alpha$+2n)\cite{Mor07} will be used in CDCC, CRC and NEB calculations.

\subsection{Optical Model calculations}\label{sec:om}

\begin{figure}[h]
\begin{minipage}{20pc}
\includegraphics[width=20pc]{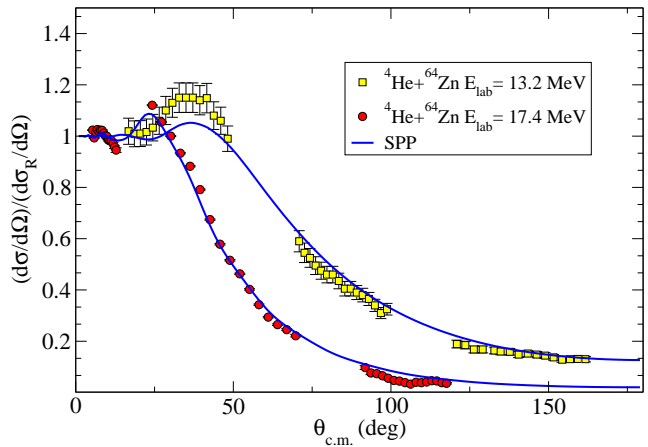}
\caption{\label{om4he}Elastic scattering angular distributions of $\alpha$+$^{64}$Zn reaction. The red circles are the experimental data at 17.4 MeV, while the yellow squares represent the experimental data at 13.2 MeV of Ref.~\cite{DiPietro03,DiPietro2004}. The solid lines are the OM calculations. }
\end{minipage}\hspace{2pc}%
\end{figure}

 The experimental data of the reaction $^{4}$He+$^{64}$Zn at 17.4~MeV  have been analyzed with the OM method, using the microscopic S{\~a}o Paulo Potential (SPP) \cite{Cha02} for the real part of the potential. For the imaginary part, the same geometry was adopted, renormalized by the factor $N_{i}$ = 0.78  from Ref.~\cite{Alv03}. In Fig.~\ref{om4he}, the calculated elastic scattering angular distributions are compared with the present experimental data and those of Refs.~\cite{DiPietro03,DiPietro2004}. Considering the uncertainty in the normalization discussed in Sec. II, the calculated elastic scattering angular distribution is in reasonable agreement with the experimental data. Therefore, from now on we will consider the SPP for the  interaction $^{4}$He+$^{64}$Zn.

\begin{table}[h]

\caption{\label{tab:pot_tot}  Parameters of the derivative imaginary Woods-Saxon potential for the $^{6}$He+$^{64}$Zn  reactions, where the reduced radius is fixed at $r_d$=1.2 fm. The volume part of the optical model is given by S{\~a}o Paulo potential.  See text for details.}

\begin{ruledtabular}
\begin{tabular}{   c  c c }
 Energy (MeV) & $W_d$ (MeV) & $a_d$(fm) \\
\hline
   9.8   & 3.23  & 1.00 \\
   13.5  & 2.87  & 1.10 \\  
   14.85 & 2.46  & 0.92 \\
   17.90  & 1.91 & 0.91 \\
\end{tabular}
\end{ruledtabular}
\end{table}

In reactions induced by halo nuclei (such as $^{6}$He \cite{Fer10}, $^{11}$Be \cite{Dipietro12} or $^{11}$Li \cite{jp_cdp2015}) it is useful to consider an optical model prescription composed by two terms: one takes into account the scattering of the core with the target by means of a volume part whereas the other represents the long-range effects produced by the nuclear halo and  is conveniently parametrized using a surface potential. Thus, the $^{6}$He+$^{64}$Zn potential is parametrized according to the following expression.
\begin{equation}
U_{\text{opt}}(R)=U_{\text{bare}}(R)+i~ W_{d}(R), \label{eq:om}
\end{equation}
where  the ``bare'' potential $U_{\text{bare}}$ is  approximated by the interaction $\alpha$+$^{64}$Zn obtained before and $W_{d}$ is the surface part represented by a derivative imaginary Woods-Saxon potential. 

\begin{figure}[h]
\begin{minipage}{20pc}
\includegraphics[width=20pc]{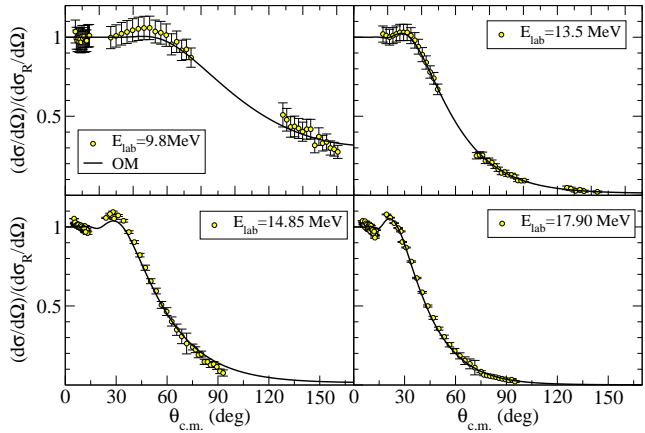}
\caption{\label{om}Experimental elastic scattering angular distributions of $^{6}$He+$^{64}$Zn reaction at the incident energies of 9.8, 13.5, 14.85 and 17.9 MeV (symbols). Optical model calculations are shown as continuous line. }
\end{minipage}\hspace{2pc}%
\end{figure}

The volume term  and the reduced radius of the surface part ($r_{d}$ = 1.2 fm) are kept fixed, while the imaginary depth ($w_{d}$) and the diffuseness ($a_{d}$) are allowed to vary in order to reproduce the elastic scattering data at the four reaction energies. The OM fits have been performed with the routine {\tt SFRESCO}, which is part of the {\tt FRESCO} coupled-channels code \cite{Thom88}.
The parameters obtained are shown in Table~\ref{tab:pot_tot} and the calculated elastic scattering angular distributions are compared with the experimental data in Fig.~\ref{om}. As  expected, a large value of the imaginary diffuseness ($a_{d}\approx$ 1 fm) is found, which can be attributed to the presence of long-range Coulomb and/or nuclear couplings (see e.g. Refs.~\cite{Fer10, DiPietro03}).

\subsection{CDCC calculations}\label{sec:cdcc}
The  $^{6}$He experimental data have also been compared with CDCC calculations. Because of  the three-body structure of $^{6}$He, we employ the four-body CDCC method. However, since it is our aim to compute also the $\alpha$ energy and angular distribution, and the calculation of these observables has not yet been  implemented in the four-body CDCC method, we have also performed  three-body CDCC calculations, assuming a two-body structure for $^{6}$He ($\alpha$+$2n$). 
Within the three-body CDCC framework, the breakup of the projectile is treated as an inelastic process, where the two valence neutrons can be excited to unbound states of the $^{4}$He-2$n$ system.  Since target excited states are not considered explicitly, the computed cross sections correspond to the elastic breakup (EBU) part defined earlier. To describe the  $^{6}$He states, the improved di-neutron model of Ref.~\cite{Mor07} has been used.   In this model, the $2n$+$^{4}$He interaction is parametrized with  a Woods-Saxon potential with a radius $R_{0}$ = 1.9~fm and diffuseness $a_{0}$ = 0.25~fm. For $\ell=0$, the depth of the potential is adjusted to give an effective two-neutron separation energy of 1.6 MeV, which was chosen to reproduce the tail and rms radius of the $2n$+$^{4}$He  wave function, as predicted by a realistic three-body calculation  of $^{6}$He.  For the $\ell$ = 2 continuum, the potential depth was adjusted to reproduce the 2$^{+}$ resonance at the excitation energy of 1.8 MeV above the ground state. 

  The CDCC calculations require also the fragment-target optical potentials. For the $^{4}$He-$^{64}$Zn interaction we considered the SPP of Subsec.~\ref{sec:om}, whereas the  $^{64}$Zn-2$n$ potential was calculated using the following single-folding model: 
\begin{equation}
 U(R)  = \int \rho(r_{nn})[ U_{n}(\vec{R} + \frac{\vec{r}_{nn}}{2}) + U_{n}(\vec{R} - \frac{\vec{r}_{nn}}{2}  )]d\vec{r}_{nn},
\end{equation}
where $\vec{R}$ is the $^{64}$Zn-2$n$ relative coordinate, $U_{n}$ is the neutron-target optical potential, which we adopted from Ref.~\cite{Kon03}, and $\rho(r_{nn})$ is the density probability along the $\vec{r}_{nn}$ coordinate calculated within the $^{6}$He three-body model of Ref.~\cite{manoli08}. The $^{6}$He continuum  was discretized using the standard binning method, including $^{4}$He-$2n$  relative angular momenta up to $\ell_{max}$ = 4 and excitation energies up to 8~MeV with respect the two-neutron separation threshold. The coupled equations were integrated numerically up to 100~fm, and for total angular momenta up to  $J$ = 80. These calculations were performed using the code {\tt FRESCO} \cite{Thom88}. 

\begin{figure}[h]
\begin{minipage}{20pc}
\includegraphics[width=20pc]{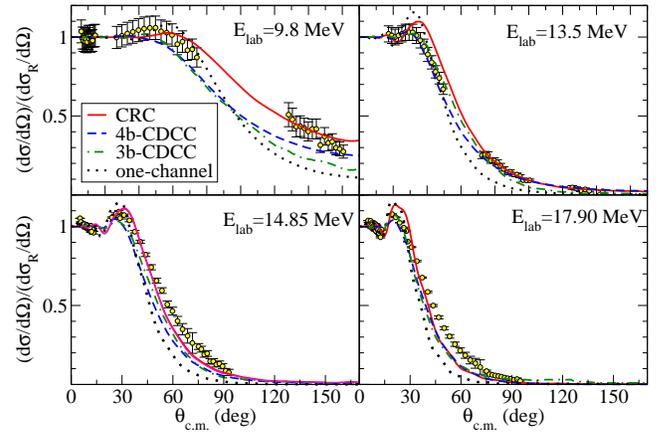}
\caption{\label{cdcc}Elastic scattering angular distributions of $^{6}$He+$^{64}$Zn reaction at the incident energies of 9.8, 13.5, 14.85 and 17.9 MeV. The dashed and dashed-dotted lines are the 4b- and 3b-CDCC calculations, respectively.  The solid line is the CRC calculation, which takes into account the coupling to the  $2n$-transfer channels (section \ref{sec:crc}). The dotted lines are the one-channel calculations, which ignore the coupling to the continuum states.}
\end{minipage}\hspace{2pc}%
\end{figure}

In Fig.~\ref{cdcc}, we compare the elastic scattering data with CDCC calculations. The solid lines are the full CDCC calculations, while the dotted lines represent the single-channel calculations, where the coupling to the continuum states is not considered. As it was found in Refs.~\cite{Fer10,Fer10plb}, the inclusion of coupling to breakup channels produces a suppression of the Coulomb nuclear interference peak. The full CDCC calculations reproduce reasonably the overall set of experimental data.


In order to assess the accuracy of the  di-neutron model of the 3b-CDCC calculations, the  elastic scattering angular distributions have been also compared with four-body CDCC calculations  using the formalism developed in Ref.~\cite{manoli08}.  To discretize the three-body continuum we use  the transformed harmonic oscillator (THO) method \cite{manoli05}.  Here we use the same structure model for the three-body system $^6$He($\alpha+ n+ n$) as in Ref.~\cite{manoli08}. The Hamiltonian includes two-body potentials plus an effective  three-body potential.  The $^6$He ground-state wave function ($j = 0^+$) needed to construct the THO basis is generated as explained in Ref.~\cite{manoli08}. 
The parameters of the three-body interaction are adjusted to reproduce the ground-state separation energy and matter radius. The calculated binding energy is 0.952 MeV and the rms radius 2.46~fm (assuming a rms radius of 1.47~fm for the $\alpha$ particle).  The fragment-target interactions were represented by optical potentials that reproduce the elastic scattering at the appropriate energy. The $n+^{64}$Zn potential was taken from the global parametrization of Koning and Delaroche \cite{Kon03}. For the $\alpha + ^{64}$Zn potential, we took the optical potential obtained in the Subsec.~\ref{sec:om}. Both Coulomb and nuclear potentials are included.
The coupled-channels equations were solved using the code {\tt FRESCO} \cite{Thom88}, with the coupling potentials supplied externally. We included in the calculation the states with angular momentum $j = 0^+$, $1^-$, and $2^+$. To get convergence  a THO basis with 86 states and truncated at the maximum energy value of 8 MeV was needed.
The coupled equations were solved up to $J = 40$ for 10 MeV and $J= 60$ for the rest of energies, and for projectile-target separations up to a matching radius of 100~fm.

 The calculated elastic angular distributions are compared with the data  in  Fig.~\ref{cdcc}. The results are very close to those obtained with the 3-body CDCC calculations, which  gives additional support to the use of the 3-body CDCC method to compute the breakup cross sections.
 
The calculated $\alpha$ cross sections, provided by the 3-body CDCC calculations, are  compared in  Fig.~\ref{dsdw} with the experimental data. A significant underestimation is observed, which is taken as an indication that the measured $\alpha$ single cross sections cannot be explained by a elastic breakup mechanism, at least in the angular range covered by the data.

In summary, although the elastic scattering can be reproduced by the CDCC calculations, showing the importance of including the coupling to breakup channels, the angular distribution of $\alpha$ particles coming from the breakup of $^{6}$He cannot be described only by a elastic breakup mechanism and, therefore, other breakup mechanisms need to be considered.

\subsection{CRC calculations}\label{sec:crc}


 The failure of the  CDCC calculations to reproduce the $\alpha$  cross sections indicates that these fragments are mostly produced via non-elastic breakup (NEB) mechanisms. The evaluation of these NEB contributions faces the difficulty that many processes can actually contribute to it, such as one- and two-neutron transfer,  complete and incomplete fusion and non-capture breakup accompanied by target excitation, as found in \cite{scuderi11}. Following previous analyses of other $^{6}$He reactions  \cite{Esc07}, an approximate way of evaluating the total inclusive cross section (i.e.\ EBU+NEB) consists in a transfer-like mechanism populating a set of doorway states of the $2n$-target system. Within the extreme di-neutron model adopted in the 3-body CDCC calculations, we consider here a two-neutron model populating a set of $2n$+$^{64}$Zn states. These states are  generated according to the procedure described in Ref.~\cite{Esc07}.  
In the present calculations, partial waves up to  $\ell_{f}$ = 5 for the $^{64}$Zn-$2n$ relative motion were considered. The states above the two-neutron breakup threshold were discretized using  2~MeV bins up to 10~MeV. For energies below the threshold, six states spaced by 2 MeV for each relative angular momentum $\ell_{f}$ were considered. The same $2n$-$^{64}$Zn and $^{4}$He-$^{64}$Zn potentials used in the CDCC calculations were considered.  In order to avoid a double counting of the effects of channel couplings, the entrance channel, $^6$He+$^{64}$Zn, has to be described by a bare potential.  For that, in this work we have used an OM potential composed by the sum of the SPP \cite{cha202} and the Coulomb dipole polarization potential \cite{And95}. In the SPP, we assumed the two-parameter Fermi (2pF) distribution with a matter diffuseness of 0.50 \cite{cha202} and 0.56 fm, for the the $^{64}$Zn and $^{6}$He matter density, respectively. In addition, and in order to simulate the fusion conditions, a short-range Woods-Saxon imaginary potential with parameters $W = 50$~MeV, $r_{0} = 1.0$~fm, $a_{0}$ = 0.1~fm was considered.



Transfer couplings were iterated beyond the first order, until convergence of the elastic and transfer observables was achieved, thus performing a CRC (coupled-reaction channels) calculation. 


In Figs.~\ref{dsdw} and \ref{dsde}, the calculated angular and energy distributions of the $^{4}$He fragments emitted in the $^6$He+$^{64}$Zn collisions are represented. A reasonable agreement between the CRC calculations and the experimental data is observed, indicating that the two-neutron transfer mechanism is a major contributor to the inclusive cross section. In the next subsection, we present further calculations for the NEB contribution based on an alternative formalism.

\subsection{Non-Elastic breakup calculations}\label{sec:nebu}

The CRC calculations presented in the previous subsection assume a transfer-like mechanism leading to a set of single-particle configurations built on top of the target ground state. These states can be interpreted as doorway states preceding more complicated configurations, involving admixtures with target excitations.  As such, the method can be regarded as an approximate way of including both EBU and NEB contributions \cite{Aus81}.


\begin{figure}[h]
\begin{minipage}{20pc}
\includegraphics[width=20pc]{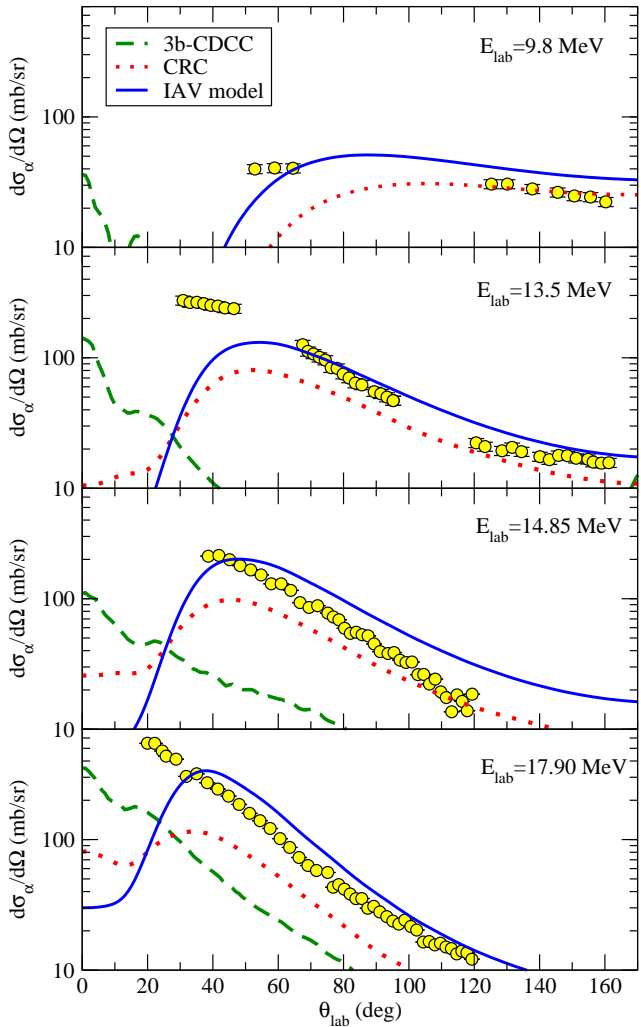}
\caption{\label{dsdw}Angular distributions of the $^{4}$He fragments, in the laboratory frame, for the reaction $^{6}$He+ $^{64}$Zn at energies of 9.8, 13.5, 14.85 and 17.9 MeV. The dotted red lines represent the CRC calculations considering the 2n transfer reaction, while the dashed green lines are the CDCC calculations, which consider the elastic breakup of the projectile. The non-elastic breakup calculations based on the IAV model are represented by solid blue lines. }
\end{minipage}\hspace{2pc}%
\end{figure}

\begin{figure}[h]
\begin{minipage}{20pc}
\includegraphics[width=20pc]{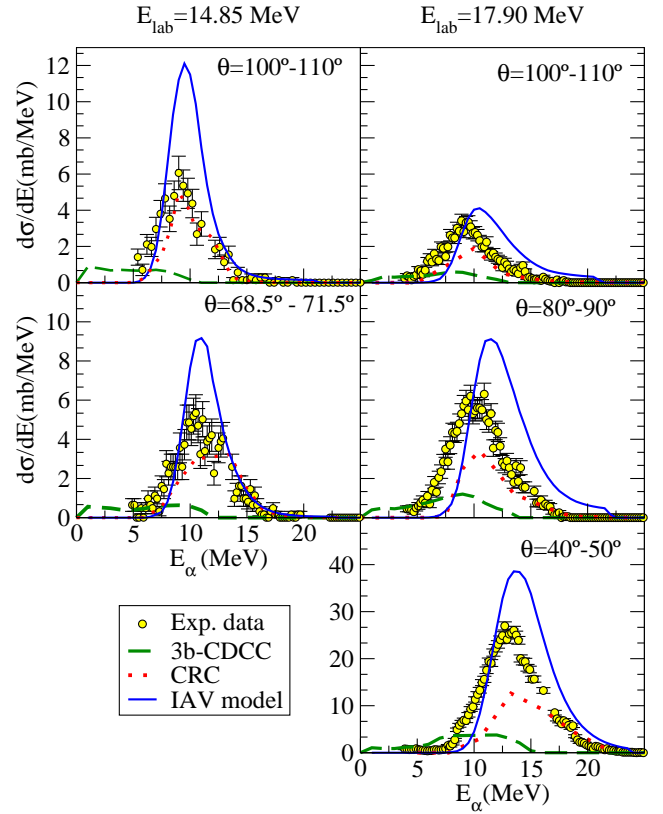}
\caption{\label{dsde}Energy distributions of the $^{4}$He fragments, in the laboratory frame, for the reaction $^{6}$He+ $^{64}$Zn at  energies of 17.9 (right panel) and 14.85 MeV (left panel) in the angular range of 40-50$^{\circ}$, 80-90$^{\circ}$ and 100-110$^{\circ}$ for the 17.9 MeV and 100-110$^{\circ}$ and the strip cover 70 $\pm$1.5$^{\circ}$ for the 14.85 MeV. The dotted red lines represent the CRC calculations considering the 2n transfer reaction, while the dashed green lines are the CDCC calculations, which consider the elastic breakup of the projectile. The non-elastic breakup calculations based on the IAV model are represented by the solid blue lines.}
\end{minipage}\hspace{2pc}%
\end{figure}

A more rigorous method to evaluate NEB cross sections was proposed in the 1980s by Ichimura, Austern and Vincent (IAV) \cite{Aus87,Ich85}, and has recently been revisited and successfully applied to several inclusive breakup reactions \cite{Jin15,Pot15,Car16}. The model is based  on a participant-spectator description of the reaction and makes use of the Feshbach projection technique. It was originally developed for processes of the form $a+A \rightarrow b +B$, where $a=b+x$ is a two-body projectile, and $B$ is any final state (bound or unbound) of the $x+A$ system. In this model, the double differential cross section, as a function of the energy and angle of the detected fragment ($b$), is given by:
\begin{equation}
\label{eq:iav_3b}
\left . \frac{d^2\sigma}{dE_b d\Omega_b} \right |_\mathrm{NEB} = -\frac{2}{\hbar v_{a}} \rho_b(E_b)  \langle \varphi_x (\vec{k}_b) | \mathrm{Im}[U_{xA}] | \varphi_x (\vec{k}_b) )\rangle   ,
\end{equation}
where $\rho_b(E_b)=k_b \mu_{b} /[(2\pi)^3\hbar^2]$,  $U_{xA}$ is the optical potential describing $x+A$ elastic scattering, and  $\varphi_x(\vec{k}_c,\vec{r}_{xA})$ is the wave function  describing the evolution of the $x$ particle after dissociating from the projectile, when the core is scattered with momentum $\vec{k}_b$ and  the target remains in its ground state. This function is obtained as a solution of the  inhomogeneous equation
\begin{equation}
\label{eq:pz_3b}
(E_x - K_x - {U}_{xA})  \varphi_x (\vec{k}_b,\vec{r}_{x}) =  \langle \vec{r}_{x}\chi_b^{(-)}(\vec{k}_b)| V_\mathrm{post}|\Psi^{3b} \rangle ,
\end{equation}
where $E_x=E-E_b$,  $\chi_b^{(-)}(\vec{k}_b,\vec{r}_{bB})$ is the distorted-wave describing the scattering of the outgoing $b$ fragment  with respect to the $B\equiv x+A$ system, obtained with some optical potential $U_{bB}$, and $V_\mathrm{post} \equiv V_{xb}+U_{bA}-U_{bB}$ is the post-form transition operator. This equation is to be solved with outgoing boundary conditions. 
In addition, for simplicity, DWBA approximation is applied to the three-body wave function, i.e.,  $| \Psi^{3b} \rangle= |\chi_a^{(+)}\phi_a \rangle$, where $\chi_a^{(+)}$ is the distorted wave describing $a+A$ elastic scattering and $\phi_a $ is the projectile ground state wave function. 

Although the model is not directly applicable to  $^6$He, due of its three-body structure,  we apply it anyway, considering one limiting scenario, in which the $\alpha$ core is the spectator and the two valence neutrons interact inelastically with the target as a whole. We present the results of these two kinds of calculations in Figs.~\ref{dsdw} and \ref{dsde}. 


The calculated NEB angular distributions reproduce rather well the shape  of the data, whereas  the magnitude is somewhat overestimated. Despite this disagreement, which might be due to the crude structure model or to the DWBA approximation itself, we can conclude that,  at the four measured energies, the non-elastic breakup of the projectile is the dominant breakup mode.

\begin{figure}[ht!]
\begin{minipage}{20pc}
\includegraphics[width=20pc]{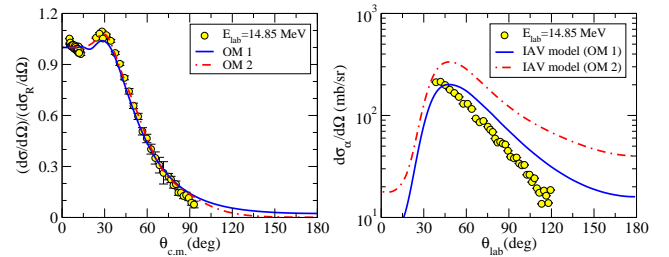}
\caption{\label{dsdw_onlyneb} (left) Elastic scattering angular distributions of $^{6}$He+$^{64}$Zn reaction at the incident energy of 14.85 MeV. (right) Angular distributions of the $^{4}$He fragments, in the laboratory frame, for the reaction $^{6}$He+ $^{64}$Zn at incident energy of 14.85 MeV.  }
\end{minipage}\hspace{2pc}%
\end{figure}

Since some ambiguities of the breakup cross-section obtained from DWBA calculations have been observed from different potentials which reproduce the entrance channel in Refs.~\cite{Leeb,Igo58,brandan97,Parey,Guyer}, we have calculated the NEB angular distributions with two different potentials at 14.85~MeV. One is the optical model potential from Table~\ref{tab:pot_tot} (OM 1), which reproduces better the experimental data in terms of $\chi^2$, and the second optical model (OM 2)  has been forced to reproduce the experimental data in the vicinity of the rainbow peak, yielding: $v$=5.71~MeV, $r$=0.99~fm, $a_{0}$=1.99~fm, $w$=0.25~MeV, $r_{i}$=1.86~fm and $a_{i}$=0.99~fm (see LHS of Fig.~\ref{dsdw_onlyneb}). The calculated breakup cross sections (shown in the RHS of Fig.~\ref{dsdw_onlyneb}), indicate a significant sensitivity  of the NEB results with respect to the entrance channel optical model potential. Therefore, improvement on  theoretical NEB model is necessary for a complete understanding of reaction involving two-neutron halo nuclei.


\section{Summary and conclusions}\label{sec:concl}
In this work we have presented new experimental data for the reaction $^{6}$He+ $^{64}$Zn at middle-target energies of 14.85 and  17.9 MeV. 
In order to understand the dynamics of the reaction, the measured elastic angular distributions, together with the data previously published in \cite{DiPietro03,DiPietro2004} have been compared with different theoretical calculations. As a start, optical model calculations based on the S{\~a}o Paulo potential, were performed. In order to reproduce the experimental data, a derivative imaginary Woods-Saxon potential with a large value of the diffuseness parameter were needed, in accord with previous findings for other $^6$He-induced reactions on medium-mass and heavy targets \cite{San08}. This is a clear indication of the presence of long-range absorption, and this is confirmed by three-body and four-body CDCC calculations, which show the need to include coupling to the continuum in order to reproduce the experimental elastic scattering data.



However, the $\alpha$ cross sections provided by these three-body CDCC calculations are found to largely underestimate the data,  indicating that the majority of observed $\alpha$ particles are not produced by an elastic breakup mechanism.  In order to pin down their origin, additional calculations have been performed and compared with the measured angular and energy distributions.

In order to investigate the role of transfer, CRC calculations were performed. They consider a two-neutron transfer mechanism populating bound and unbound states of the target nucleus. These calculations better reproduce  the shape and magnitude of the angular distributions of the $\alpha$ particles, although some underestimation is still observed. One should bear in mind, however, that this two-neutron transfer mechanism, although it may be reasonable for the $2n$-halo structure of $^6$He, it provides a very crude description of the final $2n$+$^{64}$Zn states, which depends significantly on the choice of $2n$+$^{64}$Zn energy levels.

 Finally, we consider the DWBA version of the inclusive breakup mechanism of Ichimura, Austern and Vincent (IAV). These calculations account for the non-elastic breakup processes in which the neutrons interact non-elastically with the target. Calculations based on this model reproduce very well the shape  of the $\alpha$ distributions, although they overestimate somewhat the measured cross sections.
Despite the remaining disagreement in the magnitude, which might be a consequence of the simplified two-body description of the $^6$He projectile or of the DWBA assumption itself, these calculations clearly indicate that most of the measured $\alpha$ yield  stem from non-elastic breakup mechanisms,  involving the transfer or absorption of the valence neutrons  by the target nucleus.  In order to better reproduce the breakup data,  some improvements would be necessary such as:  i) the extension of the IAV model beyond the DWBA approximation and ii) the use a three-body model of the $^{6}$He projectile in the IAV model.
 
 We outline the main conclusions of this work:
\begin{enumerate}[label=(\roman*)]
    \item $^6$He nucleus, as other halo nuclei, can be strongly polarized during the scattering. Hence, an important effect of coupling to the continuum is found in the elastic scattering, that can be described either with full 4b-CDCC or with di-neutron based 3b-CDCC calculations. These effects, which were studied in heavier targets, are now investigated in a medium-mass target $^{64}$Zn, where both Coulomb and nuclear forces play a role.
    \item The elastic break-up mechanism, as described by CDCC, fails completely to describe the yield of alpha particles coming out of the reaction. CRC calculations, based on a model in which the di-neutron is transferred to some states in the target, close to the continuum, can explain partly the alpha particle produced. However, this explanation is not completely satisfactory, as these CRC calculations depends significantly on the choice of these di-neutron-target states to which the di-neutron is transferred.
    \item The IAV non-elastic breakup mechanism for the di-neutron removal, which is completely determined by the optical potentials, is able to describe the scattering energy dependence of the alpha particle yield, the angular dependence and the energy distributions of these alpha particles. Although the mechanism overestimates the alpha particle yield (which is not surprising, given that we assume a di-neutron model), the results are relevant to correlate the elastic scattering and the inclusive break-up of halo nuclei.
\end{enumerate}

\section*{acknowledgments}
This work has been partially supported by 
the Spanish Ministerio de  Econom\'ia y Competitividad and FEDER funds (projects
 FIS2014-53448-C2-1-P, FIS2017-88410-P and FPA2016-77689-C2-1-R)  and by the European Union's Horizon 2020 research and innovation program under grant agreement No.\ 654002.

\bibliographystyle{apsrev4-1}
\bibliography{jpbiblio}

\end{document}